\documentclass[conference,10pt]{IEEEtran}

\usepackage{times}
\usepackage{graphicx}
\usepackage{amsmath}
\usepackage{amssymb}
\usepackage{color}

\newcommand{\SOC}{{\textsc{SOC}}}

\newcommand{\ls}[1]
   {\dimen0=\fontdimen6\the\font
    \lineskip=#1\dimen0
    \advance\lineskip.5\fontdimen5\the\font
    \advance\lineskip-\dimen0
    \lineskiplimit=.9\lineskip
    \baselineskip=\lineskip
    \advance\baselineskip\dimen0
    \normallineskip\lineskip
    \normallineskiplimit\lineskiplimit
    \normalbaselineskip\baselineskip
    \ignorespaces}
\begin{document}

\title{Self-* overload control for distributed web systems}

\author{\authorblockN{Novella Bartolini, Giancarlo Bongiovanni, Simone Silvestri}
\authorblockA{Department of Computer Science\\
University of Rome ``Sapienza'', Italy\\
Email: \{novella,bongiovanni,simone.silvestri\}@di.uniroma1.it}
}

\maketitle

\begin{abstract}
Unexpected increases in demand and most of all flash crowds are considered
the bane of every web application as they may cause
intolerable delays or even service unavailability.

Proper quality of service policies
must guarantee rapid reactivity and responsiveness even in such critical situations.
Previous solutions fail to meet common performance requirements
when the system has to face sudden and unpredictable surges of traffic.
Indeed they often rely on a proper setting of key parameters which  requires laborious manual tuning,
preventing  a fast adaptation of the control policies.

We contribute an original Self-* Overload Control (\SOC) policy.
This allows the system to self-configure a dynamic constraint
on the rate of admitted sessions
in order to
 respect  service level agreements and
maximize the resource utilization at the same time.
Our
policy does not require any prior information on the incoming
traffic or  manual configuration of key parameters.

We ran extensive simulations under a wide range of operating
conditions, showing that \SOC\
rapidly adapts to time varying traffic and self-optimizes the resource
utilization. It admits as many new sessions as possible in
observance of the agreements, even under intense workload variations.
We compared our algorithm to  previously proposed
approaches highlighting a more stable behavior and a better performance.
\end{abstract}


%
\IEEEpeerreviewmaketitle

\section{Introduction}
Quality of Service (QoS) management for web-based applications
is typically considered a problem of system sizing: enough resources are to be provisioned
to meet quality of service requirements under a wide range of operating conditions.
While this approach is beneficial in making the site performance satisfactory in the most
common working situations, it still leaves the site incapable to face sudden and
unexpected surges of traffic.
In these situations, in fact, it is impossible to predict the
intensity of the overload. The architecture in use,
although over-dimensioned, may not be sufficient to meet the desired QoS.
For this reason, unexpected increases of requests and most of all flash crowds are considered
the bane of every internet based application, and must be addressed in terms of performance control
rather than
capacity sizing.

Due to the ineffectiveness of static resource over-provisioning, several
alternative approaches have been proposed for overload management in web systems,
such as dynamic provisioning, dynamic content adaptation, performance degradation and admission control.
Most of the previously proposed works on this topic rely on laborious parameter tuning and
manual configuration that impede fast adaptation of the control policies.
This work is motivated by the need to formulate a fast reactive and autonomous
approach to admission control.

We contribute an original Self-* Overload Control policy (SOC) which enables some fundamental
 self-* properties such as self-configuration, self-optimization, self-protection.
In particular, the proposed system is capable of
self-configuring its component level parameters according to performance requirements.
At the same time it optimizes its own responsiveness and self-protects from overload.

The proposed policy is to be adopted by web cluster dispatching points (DP)
and does not require any modification of the
client and/or server software. DPs intercept
requests and make decisions to block or accept incoming new
sessions to meet the service level requirements detailed in a Service Level Agreement (SLA).
Decisions whether to accept or refuse new sessions are made  on the basis of a dynamically adjusted
upper limit on the admission rate. This limit  is updated and kept consistent with the
system capacity and time varying traffic behavior, measured by an
apposite self-learning, monitor module.
Such module performs an autonomous and continuous measurement activity
that is of primary importance if human supervision is to be avoided.

Our proposal is oriented to the management of web based traffic, and for this reason provides admission control at session granularity. Nevertheless, it does not require any prior
knowledge on the incoming traffic, and can be applied to non-session based traffic as well.

Unlike previous works, our approach is rapidly adaptive, and also capable to deal
with flash crowds which
are detected as soon as they arise, with a simple change detection mechanism, that permits a
 fast adaptation of the rate of decision updates.
The inter-decision time becomes increasingly shorter as traffic changes become sudden and fast,
as in presence of flash crowds.
This interval is set back to longer values when the workload conditions return to normality.

Although inspired by our previous work \cite{Noi_MASCOTS2007}, this proposal is original as it includes the anomaly detection and decision rate adaptation mechanisms necessary to perform flash crowd management.
It also provides a considerably improved measurement validation system as detailed in section \ref{sec:algorithm}.

%


We designed a synthetic traffic generator,  based on an industrial
standard benchmark SPECWEB2005, which we used to run
simulations under a wide range of operating conditions.
We compared \SOC\ to other commonly adopted approaches
showing that it outperforms the others in terms of performance and stability
even  in presence of flash crowds.
Indeed \SOC\ does not show the typical oscillations
of response time due to the over-reactive behavior of other policies.

A wide range of experiments has been conducted to test the sensitivity of the proposed solution
to the configuration
of the few startup parameters.
Experiments show that the behavior of our policy is not dependent on the initial parameter setting,
while  other  policies achieve an acceptable performance
only when perfectly tuned and in very stable scenarios.


The paper is organized as follows: in section \ref{sec:problem}
we formulate the problem of overload control in distributed web systems.
In section \ref{sec:idea} we sketch the basic actions of the proposed overload control
policy.  In section \ref{sec:algorithm} we introduce our algorithm in deeper details.
In section \ref{sec:policies} we introduce some previous approaches that we compared to ours in section \ref{sec:results}. Section
\ref{sec:related_work} outlines the state of the art of admission control in
distributed autonomic web systems
while section \ref{sec:conclusions} concludes the paper with some final remarks.

\section{The problem} \label{sec:problem}

We tackle the problem of admission control for web based services.
In this context, the user interaction with the application typically consists of a sequence
of requests forming a navigation {\em session}.
As justified by \cite{Cherkasova2002,Rom2002} we make the admission control work at session granularity.

Since the system should promptly react to traffic anomalies,
any type of solution that requires human intervention is to be excluded.
For this reason we address this problem by applying the autonomic
computing \cite{IBM-manifesto} design paradigm.

We consider a typical multi-tier architecture \cite{cola_survey,Pacifici2007}.
Each tier is composed by several replicated servers, while
a front-end dispatcher hosts the admission control and dispatch module.

Each request may involve execution at different depths in the tiered architecture.
This results in a differentiation of requests into several categories whose average processing times  may differ significantly.

The quality of service of web applications is usually regulated by a SLA.
Although our work may be applied to several formulations of SLA, when clusters of heterogeneous tiers
are considered, the most appropriate formulation is the following, as we argue in \cite{Noi_INFOSCALE2007}:

\begin{itemize}
\item{ $ RT_\texttt{SLA}^i$ }: maximum acceptable value of the  95\%-ile of the response time for requests of type $i\in\{1,2,\ldots, K\}$, where $K$ is the number of cluster tiers.
\item{$\lambda_\texttt{SLA}$}: minimum guaranteed admission rate.
If $\lambda_\texttt{in}(t)$ is the rate of incoming sessions, and
$\lambda_\texttt{adm}(t)$ is the rate of admitted sessions, this
agreement imposes that $\lambda_\texttt{adm}(t) \geq \min
\{\lambda_\texttt{in}(t), \lambda_\texttt{SLA}\}$
\item{$T_\texttt{SLA}$}: observation interval between two subsequent checks of
the satisfaction of the SLA constraints.
\end{itemize}

Meeting these quality requirements under sudden traffic variations
requires novel techniques that guarantee the necessary responsiveness.
In such cases the respect of the agreement on response time
is a challenging problem. Some other performance issues arise as well, such as
the presence of oscillatory behavior, that typically affects
some over-reacting policies, as we show in the
experimental section \ref{sec:results}.

\section{The idea} \label{sec:idea}

We designed \SOC, a session based admission control policy that self-configures a limit
on the incoming rate of new sessions.
Such limit corresponds to the maximum capacity of the system to sustain the incoming traffic without
violating the agreements on quality.
It can not be evaluated off-line because it depends on the particular traffic rate and profile
that the system has to face.

Since we do not want to rely on any prior assumption on the incoming traffic,
we introduce a monitor module
that makes the system capable to learn its capacity to face each particular
 traffic profile as it is when it comes.
For this reason we make the system measure and learn the relationship between the rate of admitted
sessions and the corresponding measure of response time.
By accurately processing raw measures, the system can ``learn''
 which is the maximum session admission rate that can be adopted in observance of the SLA requirements.
This learning activity introduces some issues such as how to time performance control,
how to aggregate measures and
how to detect changes, that will be dealt in detail in the following sections.
We just mention that as soon as a change is detected the proposed system varies the rate
of performance controls to guarantee at the same time accuracy and responsiveness.

According to our proposal the admission controller operates at the application level
of the protocol stack because session information is necessary to discriminate
which requests are to be accepted (namely requests belonging
to already ongoing sessions), and which are to be refused
(requests that imply the creation of a new session).
The cluster dispatcher can discriminate between new requests and requests belonging
to ongoing sessions because either a cookie or an http parameter are appended to the request.
This technique ensures two important benefits:
1) the admission controller can be implemented on DPs,
and does not require any modification of client and server software,
2) the dispatcher can immediately respond to non admitted requests,
sending an ``{\em I am busy\/}'' page to inform the client of the
overload situation. This avoids that the expiration of protocol time-outs affects the user perceived performance and mitigates the retrial phenomenon.

\section{Self-* Overload Control (\SOC) Policy}\label{sec:algorithm}
\SOC\ works in two modalities, namely {\em normal mode} and {\em flash crowd management mode}, switching from one to the other according to the traffic scenario being considered.
During stable load situations the timing of performance control is regularly paced at time intervals of length $T^\texttt{SOC}_\texttt{AC}$.
If a sudden change of the traffic scenario is detected,
the system enters the flash crowd management modality during which performance controls and policy updates are made more often in order to avoid a system overload.

\SOC\ provides a probabilistic admission control mechanism which
filters incoming sessions according to an adaptive rate limit $\lambda^*$.
In order to properly calculate $\lambda^*$, the monitor module takes measures to analyze the relationship between the observed Response Time (RT) and the rate of admitted sessions.
The value of $\lambda^*$ is then calculated as the highest rate that the site can support without violating the constraints on RT defined in the SLAs.
The admission control policy varies the admission probability according to a prediction of the future workload and to the estimated value of $\lambda^*$.

The behavior of our policy under normal mode is described  in figure \ref{fig:algo_normal}, while figure \ref{fig:algo_fc} describes the flash crowd management mode.

\begin{figure}
{\small
\noindent{\verb|init;|\\
}  }
{\small
\noindent{\verb|normal_mode:|\\
\verb|while ((|$t<T^\texttt{SOC}_\texttt{AC}$\verb|) AND !change_detection()){|\\
\verb|  n=n+1;|  \\
\verb|  for each session arrival {|\\
\verb|    probabilistic_admission_control;|\\
\verb|    collect_raw_measures;|\\
\verb|    }|\\
\verb|  } /* end while|\\
\verb|if change_detection()|  \\
\verb|   goto flash_crowd_mode;|\\ 
\verb|else {|\\
\verb|  update_stats;| \\
\verb|  update_curve; | \\
\verb|  update_admission_probability;|\\
\verb|  t=0;|\\
\verb|  goto normal_mode;|\\
\verb|  }|\\
}
}
\caption{Pseudo-code of SOC (normal mode)}
\label{fig:algo_normal}
\end{figure}

%
For sake of simplicity, we leave the description of the parameter
initialization (instruction \verb|init|) at the end of the algorithm description, in section \ref{sec:init}.

\medskip

\noindent
{\em Normal Mode}

At each iterative cycle $n$, the admission controller accepts new
sessions with an autonomously tuned probability $p(n)$ and collects related raw measures of response time and
 session arrival rate (more details on these phases are given in sections \ref{subsec:SAC} and \ref{subsec:raw}  ).

If no abrupt change is detected in the demand intensity, the
\verb|while| loop of the normal modality is repeated every $T_\texttt{AC}^\texttt{SOC}$ seconds.

At the end of each cycle execution, the
system processes the raw measures to calculate some statistical metrics (\verb|update_stats| instruction), such as
the mean session arrival rate  $\lambda_\texttt{in}(n)$,
the mean session admission rate $\lambda_\texttt{adm}(n)$ and
the 95\%-ile of response time $RT^i(n)$, $i \in \{1,2, \ldots, K\}$. Details on the statistics update instruction are give in section \ref{sec:stats_update}.

The execution of the \verb|update_curve| instruction is of primary importance to determine
the autonomic behavior of our policy. The system constructs the function between the observed traffic rate $\lambda_\texttt{adm}(\cdot)$ and the corresponding response time for the $K$ types of requests being
served $RT^i(\cdot)$. In paragraph \ref{sec:curve} we give complete details regarding the construction of this function by means of the statistical metrics calculated in the previous \verb|update_stats| instruction.

Before starting a new admission control cycle, the algorithm
evaluates a new limit $\lambda^*(n)$ on the
admission rate, and calculates the new session admission probability accordingly, as detailed in section \ref{sec:curve_inversion}.


While in normal mode, if a flash crowd occurs and a sudden surge in demand is detected, the system enters the flash crowd management mode.
It persists in this modality as long
as the traffic pattern keeps on varying significantly.

\medskip

\noindent
{\em Flash Crowd Management Mode}

The flash crowd management mode provides that statistical metrics are updated every time a new session arrives, thus ensuring a perfect adaptivity (\verb|update_stats| instruction).
 Although statistical metrics are updated at each session arrival, no learning mechanism is activated in flash crowd management mode, i.e. there is no \verb|update_curve| instruction, due to the high variability of the incoming traffic.

The policy returns to  normal mode only when the admission probability has been properly adapted to ensure that the instantly measured session admission rate $\lambda_\texttt{ist}$ is actually below the limit $\lambda^*$.
In  this case we can assume the unexpected surge is under control and  the policy can return to normal mode, during which
performance controls are paced at a slower and regular rate.

In the following paragraphs we discuss the details of the instructions provided in figure \ref{fig:algo_normal}
and  \ref{fig:algo_fc}.

\begin{figure}
{\small
\noindent{\verb|flash_crowd_mode:|\\
\verb|for each session arrival {|\\
\verb|  update_stats; /* calculates |$\lambda_{in}(n)$\verb|, ..., and |$S$\\
\verb|  n++;|\\
\verb|  update_admission_probability;|\\
\verb|  probabilistic_admission_control;|\\
\verb|  collect_raw_measures;|\\
\verb|  measure |$\lambda_\texttt{ist}$\verb|;|\\
\verb|  if |$\lambda_\texttt{ist}<\lambda^*$ \verb| goto normal_mode|\\
\verb|  else goto flash_crowd_mode;| \\
\verb|  }|
}}
\caption{Pseudo-code of SOC (flash crowd management mode)}
\label{fig:algo_fc}
\end{figure}

\subsection{Instruction \small{\textup{\texttt{probabilistic\_admission\_control}}} }\label{subsec:SAC}

Purpose of this instruction
 is to
limit the incoming rate to $\lambda^*(n)$ by means of a probabilistic admission control.
New sessions will be admitted with probability $p(n)$, initially set to $1$ and autonomously tuned as described in section \ref{sec:curve_inversion} on the basis of a
forecast on the session arrival rate for the next iteration.

\subsection{Instruction \small{\textup{\texttt{collect\_raw\_measures}}} }\label{subsec:raw}

This instruction
enables the collection of raw measures of the RT of all
requests belonging to the currently admitted session.
We define  $\mathcal{T}_i^n$ as the set of raw measures of  response
time  for requests of type $i, i \in
\{1,2,\ldots,K\}$ during the time interval $[t_{n}, t_{n+1})$.

\subsection{Instruction \small{\textup{\texttt{update\_stats}}}}\label{sec:stats_update}


At the execution of this instruction
raw measurements are processed to calculate some statistical parameters:
\begin{itemize}
\item{  $RT^i(n)$, that is the  $95\%$-ile of the set $\mathcal{T}_i^n$, for
$i \in \{1,2, \ldots, K \}$;}
\item{  $\lambda_\texttt{in}(n)$, that is the average incoming rate of new sessions observed during the time interval $[t_{n},
t_{n+1})$;}
\item{  $\lambda_\texttt{adm}(n)$, that is the average rate of  admitted sessions during the time interval $[t_{n}, t_{n+1})$}.
\end{itemize}

In order to ensure a proper system reactivity, all statistical metrics are calculated over the set $S$ composed by the last $\min \{\lfloor\lambda_\texttt{in}\cdot t\rfloor; \lfloor\lambda^*\cdot T_\texttt{AC}^\texttt{SOC}\rfloor\}$ admitted sessions.
In normal mode, this allows an early adaptation of the admission control probability to a possibly increased demand even if it has not yet caused the trigger of the change detection mechanism.
In flash crowd mode this ensures that the rate limit is calculated on the basis of the smallest time window that still guarantees a sufficiently numerous set of raw measures.

\subsection{Instruction \small{\textup{\texttt{update\_curve}}}}\label{sec:curve}

This instruction
provides the self-learning activity of our algorithm.
It allows the system to discover the function that relates
 the
rate of admitted sessions and the RT of each tier.

The statistics collected with the \verb|update_stats| instruction give the system the following information: during the time
interval $[t_{n}, t_{n+1})$, a rate of $\lambda_\texttt{in}(n)$ new
sessions reached the DP; only a rate of $\lambda_\texttt{adm}(n)$ of
those sessions was actually served, and the $95\%$-ile of the
response time for type $i$ requests was $RT^i(n)$.

A statistical metric calculated from samples of raw measures
as described in paragraph \ref{sec:stats_update}, taken
during a single iteration, is not reliable enough for two
reasons: first, the workload is subject to variations that may cause
transient effects; second, the number of samples may not be
sufficient to ensure an acceptable confidence level. The use of
longer inter-observation periods may  allow the collection of more
numerous samples, but it is impossible to define a sufficiently long
inter-observation period for any possible traffic situation, and the
incoming workload may vary before a sufficiently representative set
of samples is gathered. Moreover too long inter-observation period
may lead to low responsiveness of the admission policy.

The idea at the
basis of our proposal is to collect these statistics under a range
of workload levels.
At each algorithm iteration the DP acquires $K$ pairs $(\lambda_\texttt{adm}(n),RT^i(n))$ for $i=1\dots K$, where
$RT^i(n)$ is the 95\%-ile of
request RT  measured at the i-$th$ tier.

\begin{figure}[!t]
\centering
 \includegraphics[width=0.3\textwidth,angle=-90]{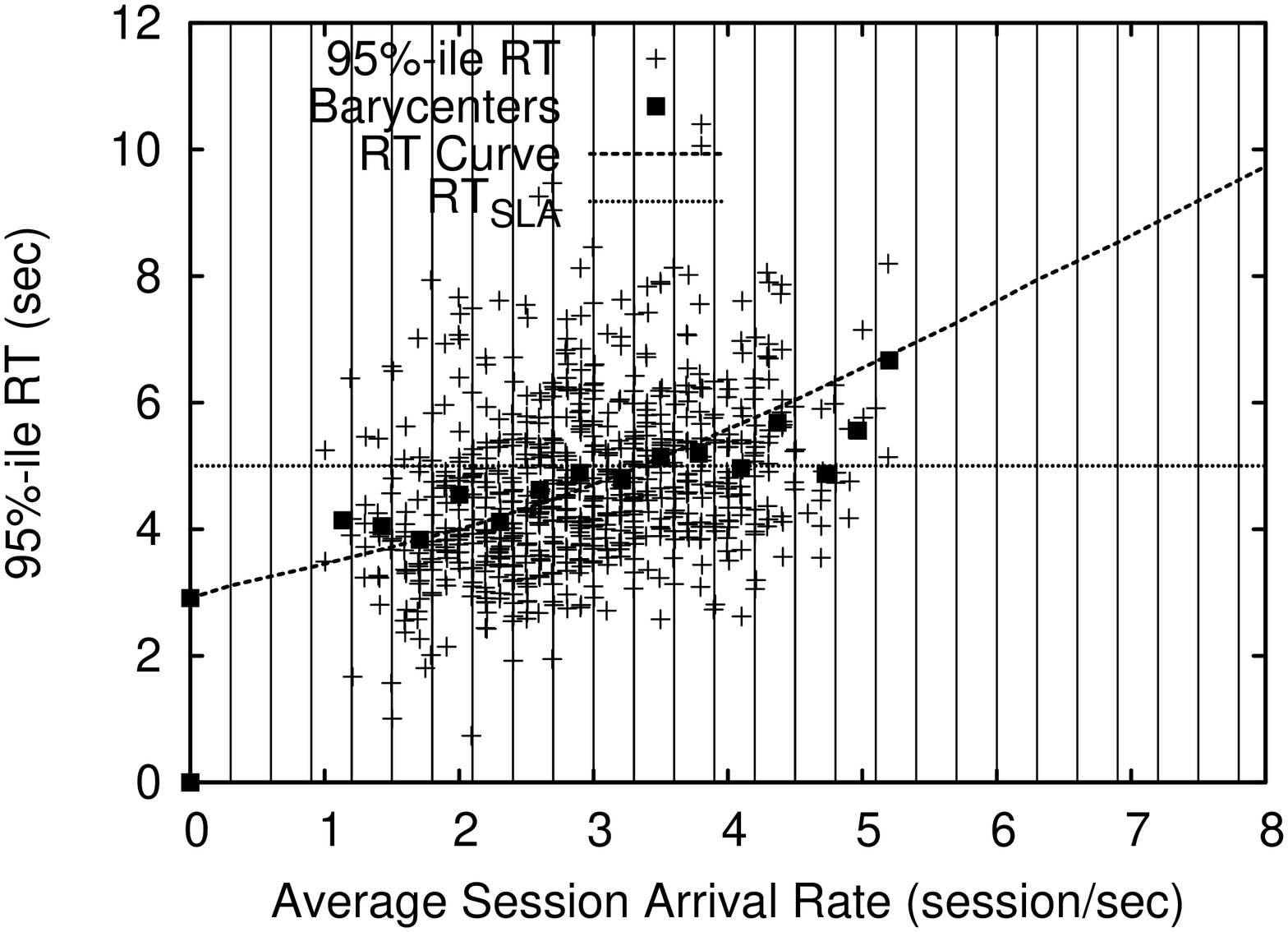}
\caption{Curve set construction, regular slice barycenters}
\label{fig:curve_set}
\end{figure}

Let us consider the set of pairs: \newline $ \mathcal{R}_i
\triangleq \left\{ (\lambda_\texttt{adm}(n),RT^i(n)), n \in \{ 0, 1,
\ldots \} \right\}$,
 where $i \in \{1,2, \ldots, K \}$, and let us partition the Cartesian plane into rectangular intervals of
length $l_\lambda$ along the $\lambda_\texttt{adm}$ axis, as shown
in figure \ref{fig:curve_set}.

For every interval $[(k-1)l_\lambda; kl_\lambda)$, with $k=1, 2,
\ldots$ we define $\mathcal{P}_k^i= \{(\lambda_\texttt{adm},
RT^i)|\lambda_\texttt{adm} \in [(k-1)l_\lambda; kl_\lambda)\} $. Then
we calculate the {\em barycenter} $B_k^i = (\lambda_k^B, RT_k^{Bi})$
of the $k$-th interval as the point with average coordinates over the set
$\mathcal{P}_k^i$.
An interval has no barycenter if $\mathcal{P}_k^i=\emptyset$.

Figure \ref{fig:curve_set} shows the collected
statistics taken at run-time at the database tier of an example
scenario. It also points out  the calculated barycenters for each interval.

Every time a new point is added to a set $\mathcal{P}_k^i$, the monitor module updates
the values of the barycenter coordinates, standard deviation and cardinality of the
set being modified.
 Notice that the
update of such values is performed for only one set at a time
(set that have not been modified do not require statistic updates) and is incrementally calculated with respect to a synthetic statistical representation. Such representation permits to  avoid computational and storage costs that would be afforded if all the pairs had to be considered.

Barycenters calculated with a standard error higher than 20\% are discarded while the others are considered sufficiently reliable and are included in corresponding lists $L^i$, where $i \in \{1,2, \ldots, K\}$. The elements of such lists are ordered on the basis of the first coordinate $\lambda_\texttt{adm}$.

\begin{figure}[!t]
 \centering
\includegraphics[width = 0.3\textwidth,angle=-90]{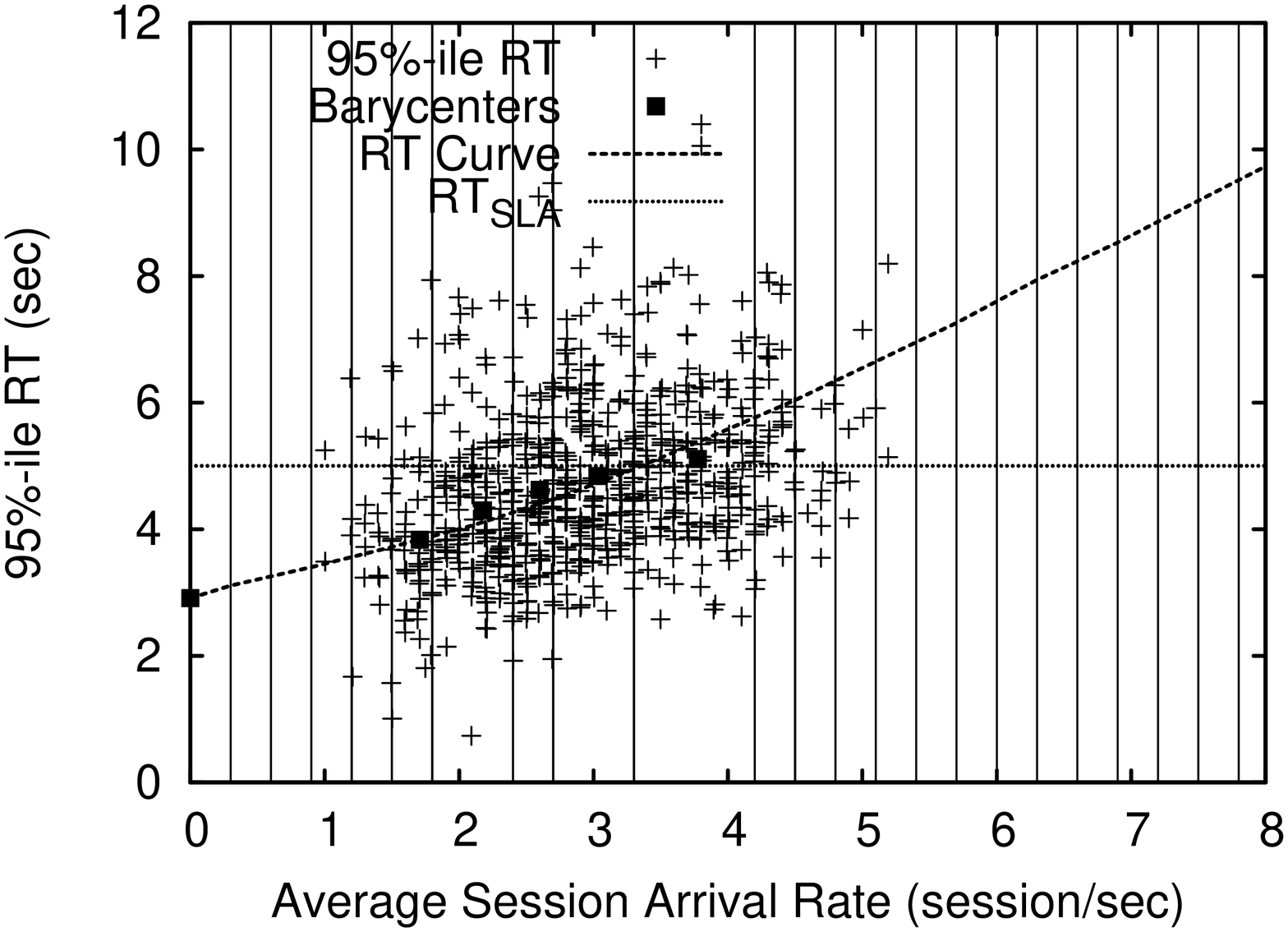}
\caption{Curve set construction, aggregated slice barycenters}
\label{fig:aggregated_regressogram}
\end{figure}

Since we know that the relation between $\lambda_\texttt{adm}$  and $RT^i$ is monotonically not decreasing, we can assume that if two subsequent barycenters do not satisfy this basic monotonicity property, the corresponding slices can be aggregated to improve the measure reliability.
For this reason, if $L^i$ contains two adjacent points which do not correspond to growing values of  $RT^i$, the sets of statistics related to the corresponding intervals are aggregated and $L^i$ is updated until it contains a list of pairs in growing order in both the coordinates, as shown in figure
\ref{fig:aggregated_regressogram}.
Notice that this procedure permits a further validation of the measures, beyond the already performed test on the standard error value.


After few aggregations, the list $L^i$ contains an ordered set of pairs which can be linearly interpolated to obtain an estimate of the function that relates $\lambda_\texttt{adm}$ and $RT^i$.
Thanks to the frequent updates, this list is a highly dynamic structure, that continuously adapts
itself to changing workload situations.

The linear interpolation of the points in $L^i$ permits to
forecast the response time corresponding to any possible workload
rate.

Notice that the use of common regression techniques as an alternative to linear interpolation  is unadvised, because it would require a prior assumption on the type of functions being parametrized for the regression.
Experiments we conducted on different traffic profiles (e.g. by using SPECWEB2005 \cite{specweb2005} and TPC-W \cite{tpc-w_1} oriented traffic generators) show that, apart from  monotonicity, no other structural property is generally valid for all the possible traffic scenarios. This would make it difficult to choose the type of regression (polynomial, exponential, power law) to use.

 \subsection{Instruction \small{\textup{\texttt{update\_admission\_probability}}}} \label{sec:curve_inversion}

The self-constructed set  $L^i$ described in paragraph  \ref{sec:curve} is linearly interpolated to obtain an estimate of the function
 ${f^i}(\cdot)$ that relates $\lambda_\texttt{adm}$ and the
 95\%-ile of response time measured at the $i$-th tier.
Such function is then used to evaluate the
highest
session admission rate $\lambda^*$ that can be adopted to remain under the response time constraints defined in the SLA.

Thanks to this estimation, the DP can configure
the session admission probability according to a forecast of the
incoming workload.

The algorithm is based on a prediction of the
session arrival rate $\hat{\lambda}_\texttt{in}(n)$ for the next
iteration interval $[t_{n}, t_{n+1})$.
It assumes that an esteem of the current session arrival rate $\hat{\lambda}_\texttt{in}(n)$
can be based on the  incoming
session rate $\lambda_\texttt{in}(n-1)$ observed during the previous
interval $[t_{n-1}, t_n)$, that is, $\hat{\lambda}_\texttt{in}(n)$ =
$\lambda_\texttt{in}(n-1)$. The algorithm is sufficiently robust to
possibly  false predictions, as they  will be corrected at the next
iteration, making use of  updated statistics.

New sessions will be admitted with probability $p(n)=\min \{1,
\lambda^*(n)/\hat{\lambda}_\texttt{in}(n)\}$. This way, if the
incoming rate of new sessions in the present time interval is the
same observed in the previous, the  upper limit on the total
incoming rate of new sessions is met.

The
on-line self-tuning of the admission probability has several
benefits. On the one hand the highest possible rate of incoming
sessions is admitted, optimizing the system utilization. On the other hand it
prevents the system from overload, by quickly reducing the admission
probability as the traffic grows.

 The execution of this instruction starts with a test to verify the validity of
the rate limit $\lambda^*(n-1)$  adopted in the previous time
interval. To this extent we define two types of error in the
evaluation of $\lambda^*(n-1)$:
\begin{itemize}
\item{$ \textrm{ } error^-$: The system admitted new sessions with probability $p(n-1)$
but the incoming rate was unexpectedly greater than $\lambda^*(n-1)$.
In such a situation, if the rate limit was properly estimated, some
SLA limits should have been violated. In  this erroneous situation,
although the rate limit was exceeded, the SLA limits were not
violated. The occurrence of this error depends on a possible
underestimation of the rate limit $\lambda^*(n-1)$. More formally, if
$\lambda_\texttt{adm}(n-1) \geq \lambda^*(n-1)$ AND $\forall
i\in{1,2,\ldots, K}$ $RT^i <  RT_\texttt{SLA}^i$) then
$error^-=true$.}

\item{$\textrm{ } error^+$:  The system admitted new sessions with probability $p(n-1)$ and, as expected, the incoming rate was lower than $\lambda^*(n-1)$. In such a situation, if the rate limit was properly estimated, there should not be any violation of the agreements.
In this erroneous situation, although the rate limit was not
exceeded, a violation of at least one of the SLA limits was
observed. The occurrence of this error reveals a possible
overestimation of the rate limit $\lambda^*(n-1)$. More formally, if
($\lambda_\texttt{adm}(n-1) \leq \lambda^*(n-1)$ AND $\exists
i\in\{1,2,\ldots, K\}$ s.t. $RT^i >
 RT_\texttt{SLA}^i$) then $error^+=true$.}
\end{itemize}

If none of these errors occurred, the upper limit on the rate of
admitted sessions was properly set and there is no need to change
the value of the rate limit. Therefore, in absence of errors,
 $\lambda^*(n)=\lambda^*(n-1)$.

If otherwise one of these two types of error has occurred the value
of $\lambda^*(n-1) $ needs to be updated. To this purpose the set
$L^i$ is linearly interpoled and the resulting function $f^i(\cdot)$
is inverted in correspondence to the value of the SLA limit on the
95\%-ile of the response time $RT_\texttt{SLA}^i$. The function
$f^i(\cdot)$
 crosses
the line $t=RT_\texttt{SLA}^i$ in a point $P_i^*=(\lambda_i^*(n),
RT_\texttt{SLA}^i)$, whose first coordinate, $\lambda_i^*(n)$, is
the estimated optimal session admission rate for the $i$-$th$ tier.

 To guarantee the fulfillment of the SLA on each
tier, the optimal admission rate for the next round is set as
follows: $ \lambda^*(n) = \min_{i =1,\dots,K}\lambda_i^*(n) $.

Notice that at the startup, $L^i$ may contain only one point (the benchmark point described in paragraph \ref{sec:curve}) or several points
 located below the SLA constraint.
In the first case, the admission probability $p(n)$ is set to 1.
In the second case the linear interpolation between the extreme two points in $L^i$
is prolonged until it crosses the SLA constraint.

\subsection{Function \small{\textup{\texttt{change\_detection()}}}} \label{sec:anomaly}

This mechanism consists of two joint controls and triggers only if both of them give a positive result:
1) the number of sessions admitted during the current execution cycle (we call it $N$) exceeds the expectations for a single cycle, that is  $(N>\lambda^* \cdot T_\texttt{AC}^\texttt{SOC})$;
2) the current admission rate exceeds the limit $\lambda^*$ by $k$ times the
measured standard deviation of the admitted rate,
that is $( (N/t)>(\lambda^*+k \cdot \sigma_\lambda)   )    $, where
$t$ is the time elapsed from the start of the current iteration.
Notice that the value of  $\sigma_\lambda$ is calculated at run-time by measuring the standard deviation of the admitted rate  $\lambda_\texttt{adm}$ in situations where
$\lambda_\texttt{in}$  greater than $\lambda^*$.
It measures the intensity of the inherent variability of the admitted rate $\lambda_\texttt{adm}$, that cannot be filtered by a probabilistic admission control.

The pseudo-code of the change detection mechanism is described in figure
\ref{fig:change_detection}.

\begin{figure}
{\small
\noindent{
\verb|Boolean change_detection() {|\\
\verb|  if (|$(N>\lambda^* \cdot T_\texttt{AC}^\texttt{SOC})$
\verb| AND |
$( (N/t)>(\lambda^*+k \cdot \sigma_\lambda)   )    $\verb|)|\\
\verb|  return TRUE;|\\
\verb|  else return FALSE;|\\
\verb|  }|
}}
\caption{Pseudo-code of the change detection mechanism}
\label{fig:change_detection}
\end{figure}

\subsection{Instruction \small{\textup{\texttt{Init}}}}\label{sec:init}

The autonomic behavior of our algorithm makes the system capable of
adapting itself to changing traffic conditions when prior knowledge
of the traffic parameters is useless or even misleading. For this
reason the initial setting of the system parameters is not of
primary importance.
As initial setting of our algorithm we use $n=0$, $\lambda^*(0)=\lambda_\texttt{SLA}$ and $p(n)=1$.
As initial setting of the curve construction phase, we insert  the point $P_\texttt{bench}^i=(0,RT_\texttt{bench}^i)$ in $L^i$,  representing
the lower bound on the $95\%$-ile of the response times of type $i$ requests.
This point is the $95\%$-ile of response time
measured at the $i$-th tier, when  the system is in a completely idle state, that is
when $\lambda_\texttt{adm}\widetilde{=}0$.

In order to calculate the
average response time in such situation we use an offline benchmark,
obtaining the points $P_\texttt{bench}^i=(0,RT_\texttt{bench}^i)$,
$i \in \{1,2, \ldots, K\}$.

The proper setting of the points $P^i_\texttt{bench}$ with value
 $P_\texttt{bench}^i=(0,RT_\texttt{bench}^i)$ as detailed in section \ref{sec:curve},
is  not a key point in the algorithm, since it can be
substituted with the origin $O=(0,0)$, with no impact but a little
difference in the time to converge to a stable choice of
$\lambda^*(n)$. The use of this point in the interpolation of the
curve obtained from the set $L^i$ is in fact limited
to the first executions of the instruction
\verb|update_curve|, when too few reliable points are
available.

\section{Other admission control strategies}\label{sec:policies}
In this section we describe other previously proposed QoS policies
to make performance comparisons. These policies can be formulated in
many variants depending on the considered performance objective. We
limit our analysis to the optimization of response time which is
strictly related to the user perceived quality.


\subsection{Threshold Based Admission Control}\label{sec:on-off}
Fixed threshold policies have been proposed in many fields of
computer science, and in particular for web applications with
several variants \cite{Cherkasova2002,Elnikety2004,Noi_IPSDC}.

According to the Threshold Based Admission Control (TBAC) policy,
the DP makes periodic evaluations of the $95\%$-ile of  response
time of each tier, every $T_\texttt{AC}^{\texttt{TBAC}}$ seconds. If there is at
least one tier for which the $95\%$-ile of  response
time  exceeds a threshold
$RT^{\texttt{TBAC}}$, the DP rejects new sessions and
only accepts requests that belong to ongoing sessions. On the
contrary, if the value of the $95\%$-ile of response
time at each tier is lower than $RT^{\texttt{TBAC}}$, all new sessions are
accepted for the next $T_\texttt{AC}^\texttt{TBAC}$ seconds.


This policy, like all threshold based policies, implies a typical
on/off behavior of the admission controller. This causes
unacceptable oscillations of response time.
Furthermore, its performance  depends on a proper
parameter setting (i.e. the choice of the threshold
$RT^{\texttt{TBAC}}$ and of the period between two
succeeding decisions $T_\texttt{AC}^{\texttt{TBAC}}$), and for this reason
it cannot be used in traffic scenarios characterized by highly
variable workloads.


\subsection{Probabilistic Admission Control}\label{sec:probabilistic_scheme}
Probabilistic Admission Control (PAC) is a well known technique
in control theory, commonly used when oscillations are to be
avoided. This policy was proposed for Internet services in
\cite{Xu2004}, while a similar version was also proposed for web systems in
\cite{Aweya2002}. According to this policy, a new session is
admitted with a certain probability, whose value depends on the
measured response time.

The DP evaluates, every $T_\texttt{AC}^{\texttt{PAC}}$ seconds, the
response time of each tier. It compares the measured response times
with two thresholds, $RT_\texttt{low}^{\texttt{PAC}}$ and
$RT_\texttt{high}^{\texttt{PAC}}$.
The acceptance probability for the $i$-$th$ tier is a piece-wise
linear function of the measured $95\%$-ile of the response time
$r_i$, and has the following formulation:
\begin{equation}\label{eq:PAC}
p(r_i) \triangleq \left\{ \begin{array}{cl}
1 & \textrm{if $r_i \leq RT_\texttt{low}^{\texttt{PAC}}$}\\
\frac{RT_\texttt{high}^{\texttt{PAC}} - r_i}{RT_\texttt{high}^{\texttt{PAC}}-RT_\texttt{low}^{\texttt{PAC}}} & \textrm{if $RT_\texttt{low}^{\texttt{PAC}} < r_i  \leq RT_\texttt{high}^{\texttt{PAC}}$}  \\
0 &  \textrm{if $r_i >  RT_\texttt{high}^{\texttt{PAC}}$} \\
\end{array} \right.
\end{equation}
Then the session admission probability for the next round is given
by: $p = \min_{i=1,\dots,K}p(r_i)$

Notice that the two threshold values,
$RT_\texttt{high}^{\texttt{PAC}}$ and
$RT_\texttt{low}^{\texttt{PAC}}$, that characterize this policy, are
arbitrarily set offline independently of the observed incoming
session rate and of the inter-observation period
$T_\texttt{AC}^{\texttt{PAC}}$. Therefore, the performance of this policy
is dependent on a proper tuning of these parameters, as we show in
section \ref{sec:results}.

\section{Simulation Results}\label{sec:results}
In order to make performance comparisons
among the different policies and to investigate the
flash crowd management capabilities of SOC, we developed
a simulator on the basis of the OPNET modeler software \cite{opnet}.

In our experimental setting, we assume that the interarrival time of
new sessions follows a
negative exponential distribution. The
interarrival time of requests belonging to the same session is more
complex.
In order to have a realistic traffic generator, we used the phase
model of an industrial standard benchmark: SPECWEB2005
\cite{specweb2005}.
We refer
to \cite{specweb2005} for a detailed description of the state model
and of the functionalities of each phase.

Upon reception of a response, the next request is sent after a
think time interval $T_\texttt{think}$ spent by the  user analyzing the
received web page. Our model of $T_\texttt{think}$ is based on TPC-W
\cite{tpc-w_1,tpc-w_2} and on other works in the area of web traffic
analysis such as \cite{Weinreich2006}. As in the TPC-W model, we
assume an exponential distribution of think times with a
lower bound of 1 sec. Therefore $T_\texttt{think}=
\max\{-\log(r)\mu,1\}$ where $r$ is uniformly distributed in the
interval [0,1] and $\mu = 10$ sec. To model a realistic user
behavior, we also introduce a timeout
to
represent the maximum response time tolerable by the users. After
that a request has been sent, if the timeout expires before the
reception of the response, the client abandons the system.

We assume each phase of the session
state model can be mapped onto a specific tier of a $3$-tier
cluster.
We use an approximate estimate of the average processing times of
the different tiers on the basis of the experiments detailed in
\cite{Elnikety2004}. We assume each session phase requires an
exponentially distributed execution time set as follows: average execution time of pure
http requests is 0.001 sec, while for servlet request is 0.01 sec
and for database requests is 1 sec.

For sake of brevity, we conduct our analysis on the database tier
which  is the  bottleneck of the
architecture considered in these simulations. Thus, for simplicity,
we indicate the limit on the database response time, defined in the
SLA, as $RT_\texttt{SLA}$. All the experiments of this section are
conducted with 20 application servers, a client timeout of 8 sec.
and $RT_\texttt{SLA} = 5$sec.

The fixed threshold $T_{\texttt{AC}}^{\texttt{TBAC}}$ of the TBAC
policy is always set in agreement with the SLA constraints on the
$95\%$-ile of database response time, therefore
 $T_{\texttt{AC}}^{\texttt{TBAC}}$ = $RT_{\texttt{SLA}}$.
The thresholds of the PAC policy are defined as follows:
$T_\texttt{low}^{\texttt{PAC}}$ = 3 sec and
$T_\texttt{high}^{\texttt{PAC}}$ = $RT_{\texttt{SLA}}$, in agreement
with the SLA constraints.

\begin{figure}[!h]
\centering\includegraphics[width = 0.25\textwidth, angle=-90]{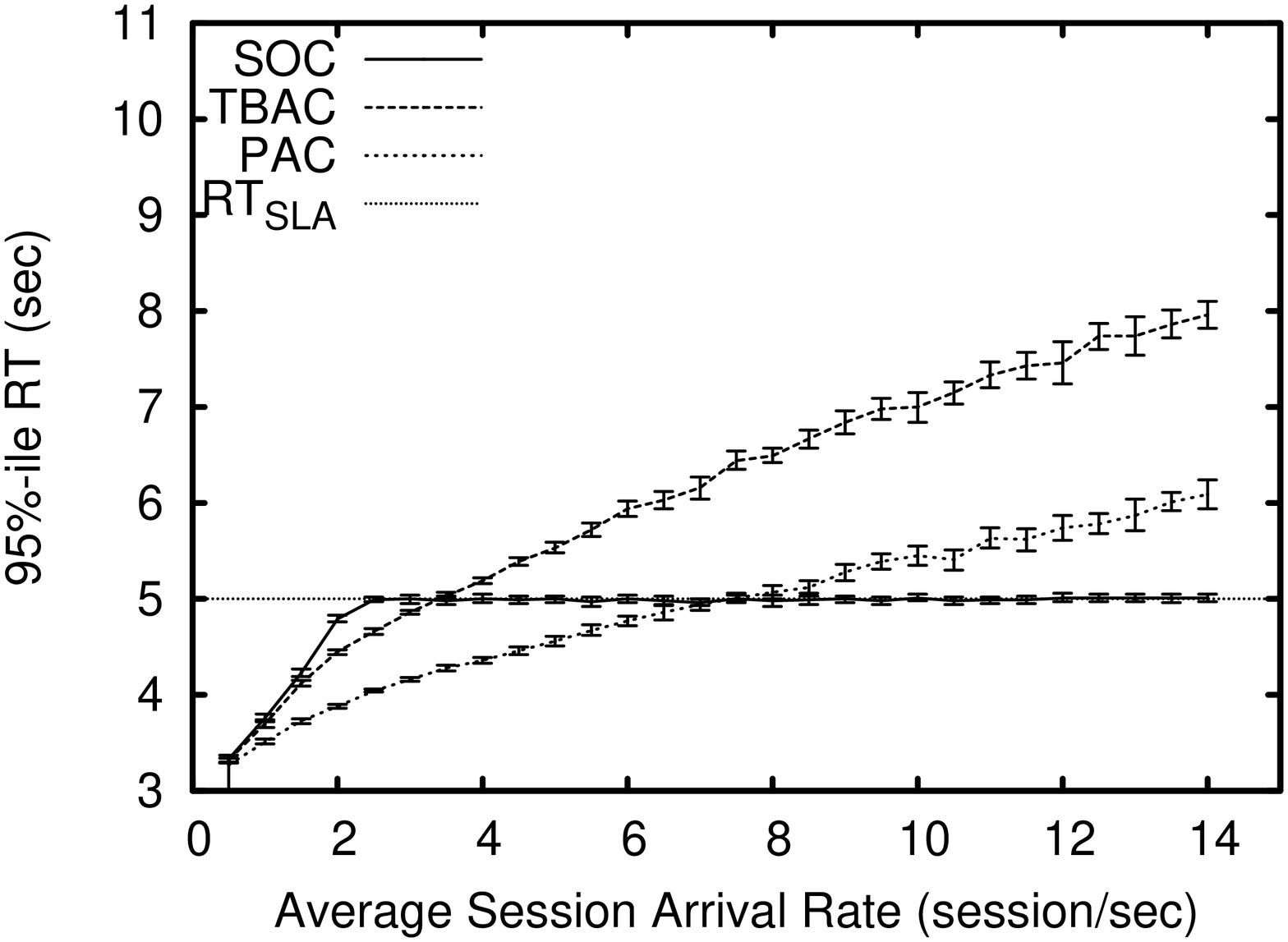}
\caption{95\%-ile of database RT} \label{fig:perc_lambda}
\end{figure}

\begin{figure}[!h]
\centering\includegraphics[width = 0.25\textwidth, angle=-90]{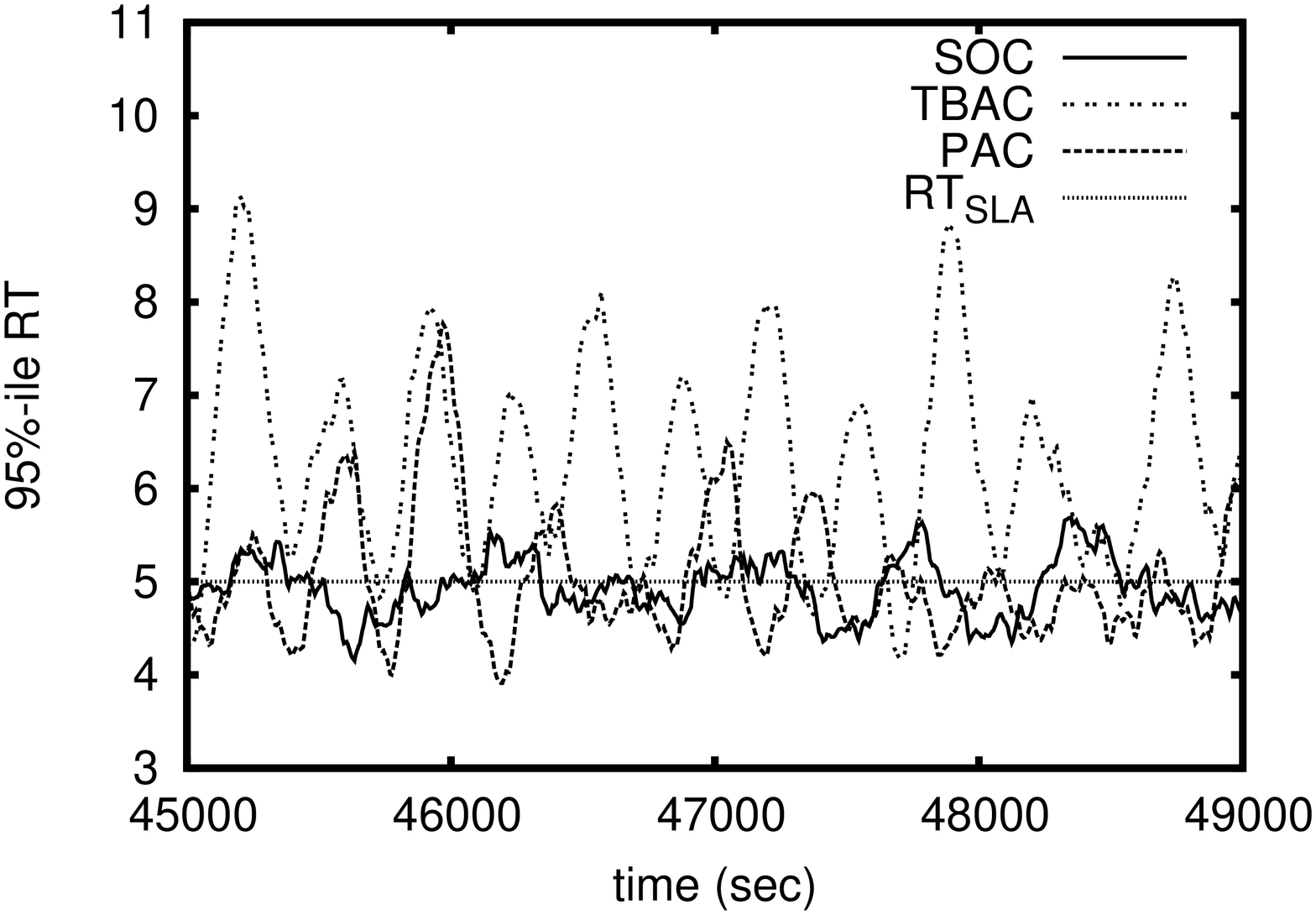}
\caption{Oscillations of 95\%-ile of database RT} \label{fig:osc_rt}
\end{figure}

A first set of experiments (figures \ref{fig:perc_lambda} and
\ref{fig:osc_rt}) shows how SOC outperforms the TBAC and PAC
policies, in terms of both
performance and stability.

Figure \ref{fig:perc_lambda} highligths  the adaptive behavior of SOC. On the one
hand, when the traffic load is
high, SOC finds the suitable session arrival rate and admits
as many sessions as possible to remain under the SLA limits. On the
other hand, when the traffic is low, it accepts almost all incoming
sessions.

Unlike SOC, other non adaptive policies, such as TBAC and PAC, typically under-utilize the
system resources in low workload conditions, and violate the QoS
agreements when the workload is high.

SOC ouperforms TBAC and PAC also in terms of stability.
As figure \ref{fig:osc_rt}
points out,  TBAC  shows an evident oscillatory behavior due to its
on/off nature while PAC has an over-reacting behavior in many situations. SOC, instead, shows a more stable response time. The self-learning activity allows to
build a reliable knowledge of the system capacity with respect
to the incoming traffic that is used to derive a good and stable
estimation of the optimal admission rate.

\begin{figure}[h]
\centering\includegraphics[width = 0.25\textwidth, angle=-90]{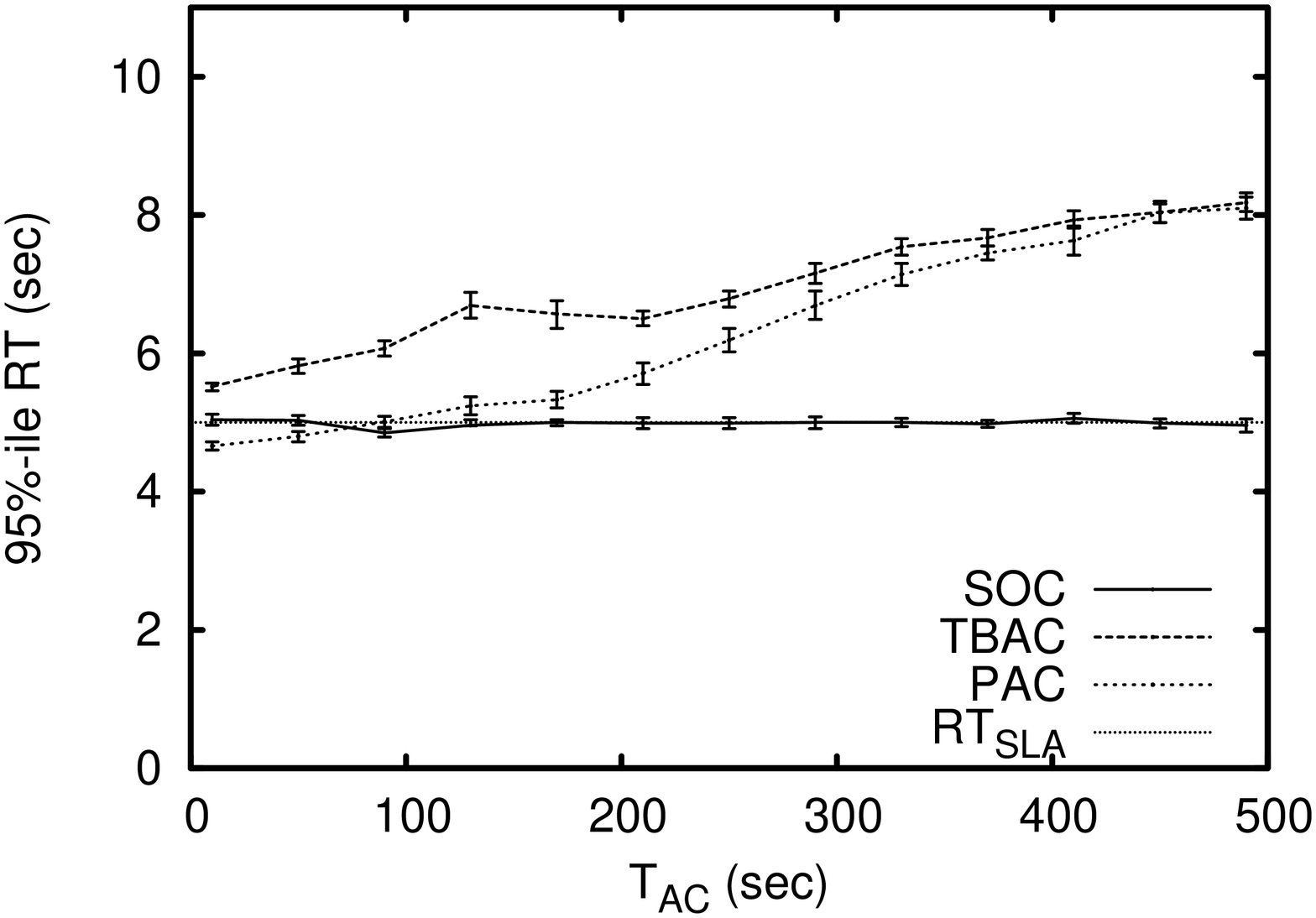}
\caption{95\%-ile of database RT} \label{fig:perc_periodo}
\end{figure}

\begin{figure}[!h]
\centering\includegraphics[width = 0.25\textwidth, angle=-90]{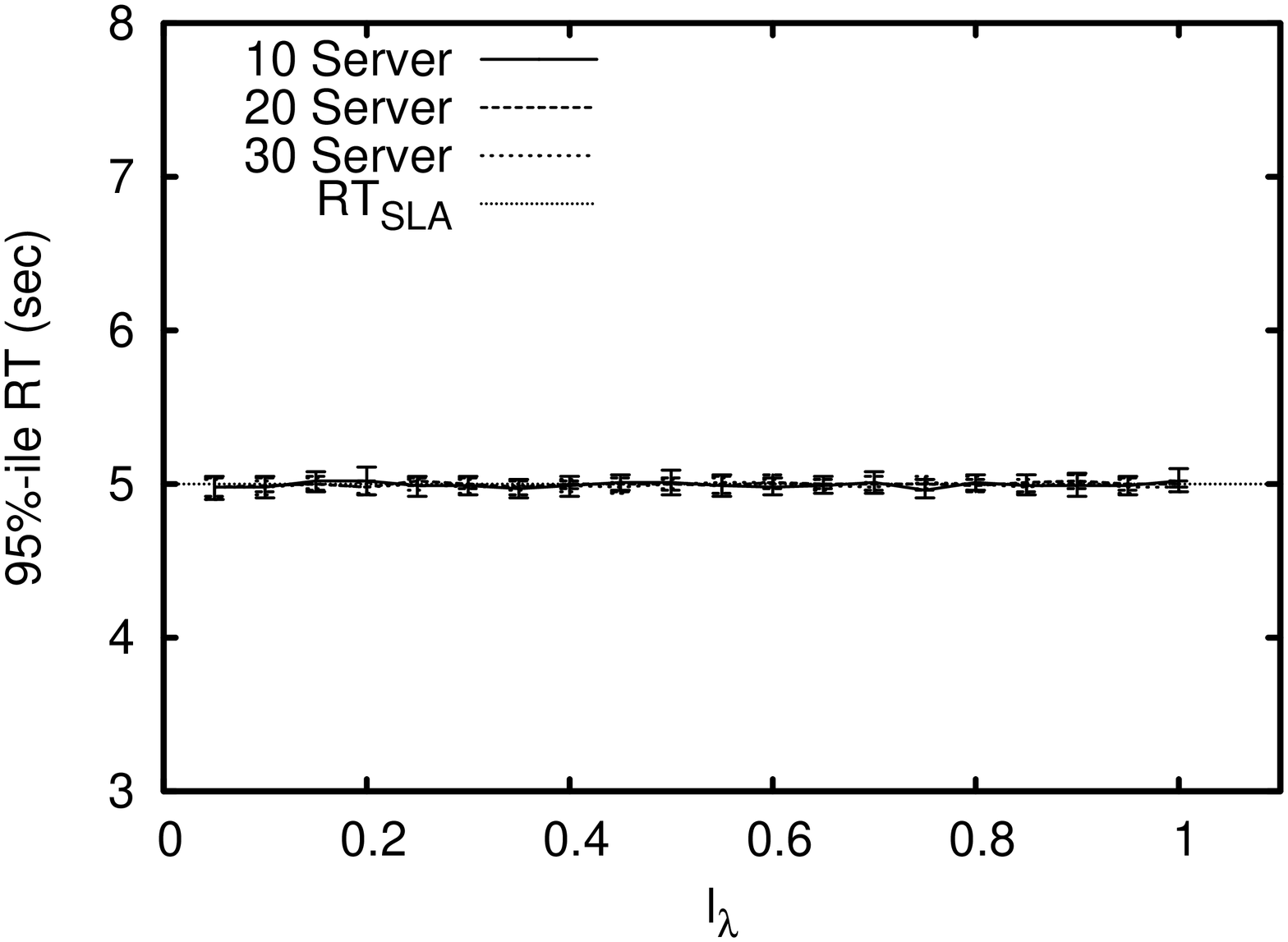}
\caption{95\%-ile of database RT} \label{fig:perc_K}
\end{figure}

\begin{figure}[!h]
\centering\includegraphics[width =0.25\textwidth, angle=-90]{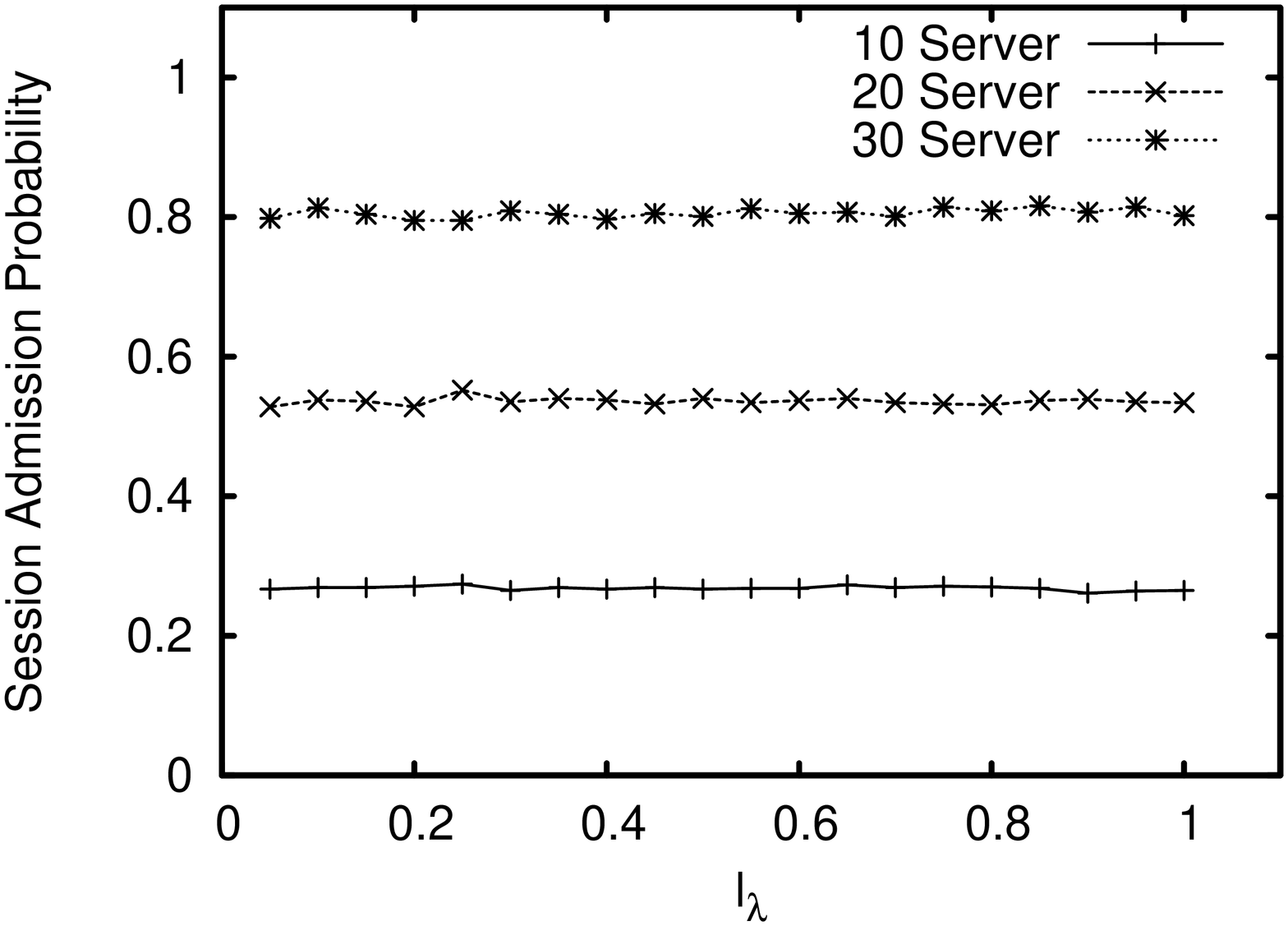}
\caption{Session admission probability}
\label{fig:p_adm_K}
\end{figure}

With the following experiments we want to show that
although SOC is based on the off-line configuration of some
parameters, (in particular $T_\texttt{AC}$ and $l_\lambda$), this does not harm its autonomy.
In fact the experiments detailed in figures
\ref{fig:perc_periodo}, \ref{fig:perc_K} and
\ref{fig:p_adm_K}
 show that the policy behavior is insensitive to the particular setting of those parameters.
These experiments were conducted with slow varying traffic scenarios. In this experimental
setting,
the particular choice of $T_\texttt{AC}$ does not influence the policy performance. Furthermore although small values of $T_\texttt{AC}$ may cause frequent triggers of the change detection mechanism (due to false positive results of the tests described in section \ref{sec:anomaly}), these triggers only cause more mode switches, without significant impact on performance (figure
\ref{fig:perc_periodo}).

Similarly the choice of the interval size $l_\lambda$ that defines the curve construction and determines the occurrence of aggregation of measurement sample sets, does not affect SOC performance.
Both response time and admission probability are stable (figures  \ref{fig:perc_K} and
\ref{fig:p_adm_K})
even when $l_\lambda$ varies significantly.

Given the slow varying traffic scenario that characterizes the experimental setting of the previous experiments,
we did not
show any performance comparison with the AACA policy that we introduced in \cite{Noi_MASCOTS2007}.
In fact in this scenario the performance of SOC is only marginally better than AACA, and the lines in the figures would have overlapped each other in many cases.

In the following experiments we studied the performance of
SOC with and without activating the change detection and flash crowd management capability described in section \ref{sec:algorithm}.

\begin{figure}[!h]
\centering\includegraphics[width = 0.25\textwidth, angle=-90]{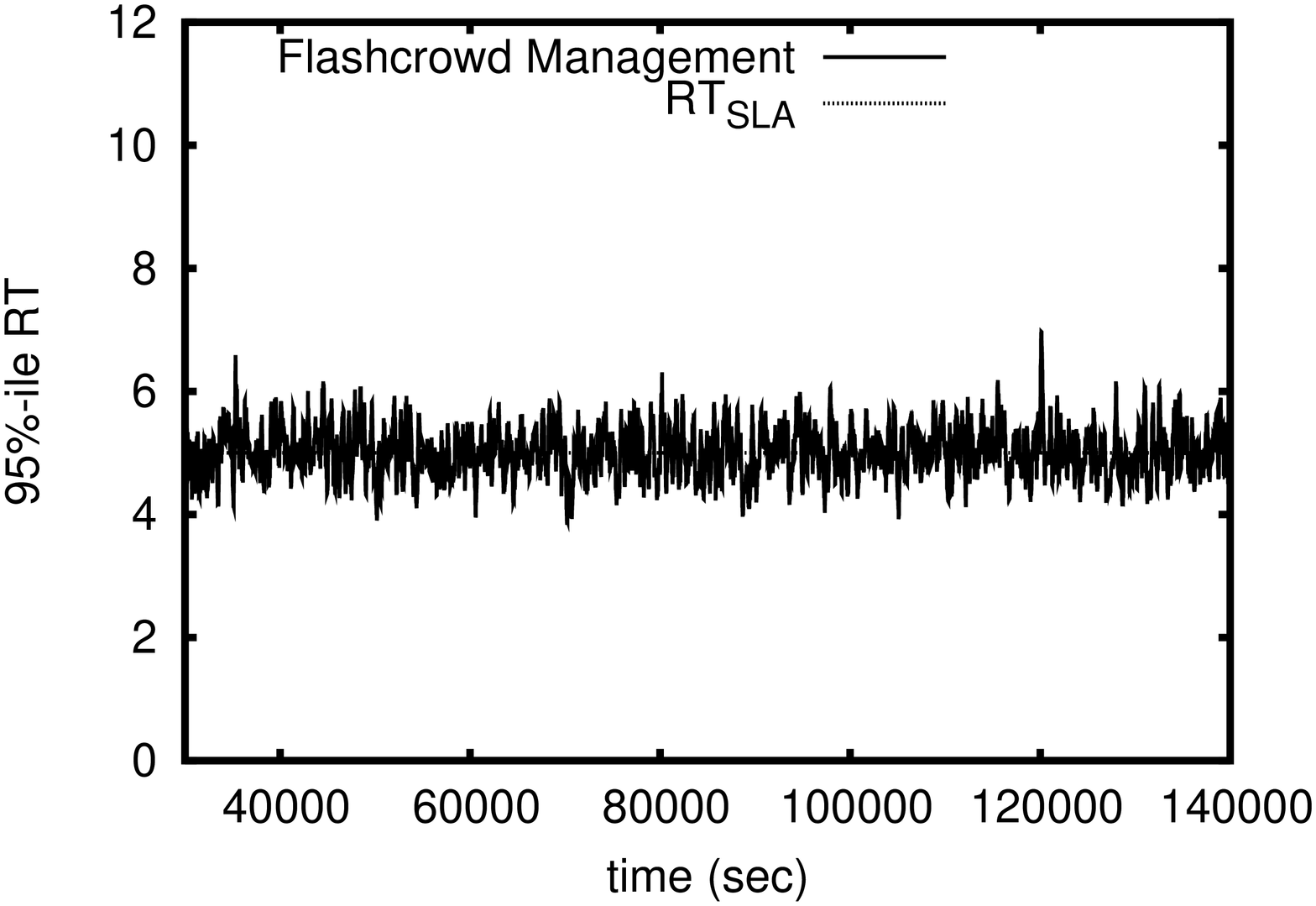}
\caption{95\%-ile RT (FCM)} \label{fig:rt_FCM}
\end{figure}

\begin{figure}
\centering\includegraphics[width = 0.25\textwidth, angle=-90]{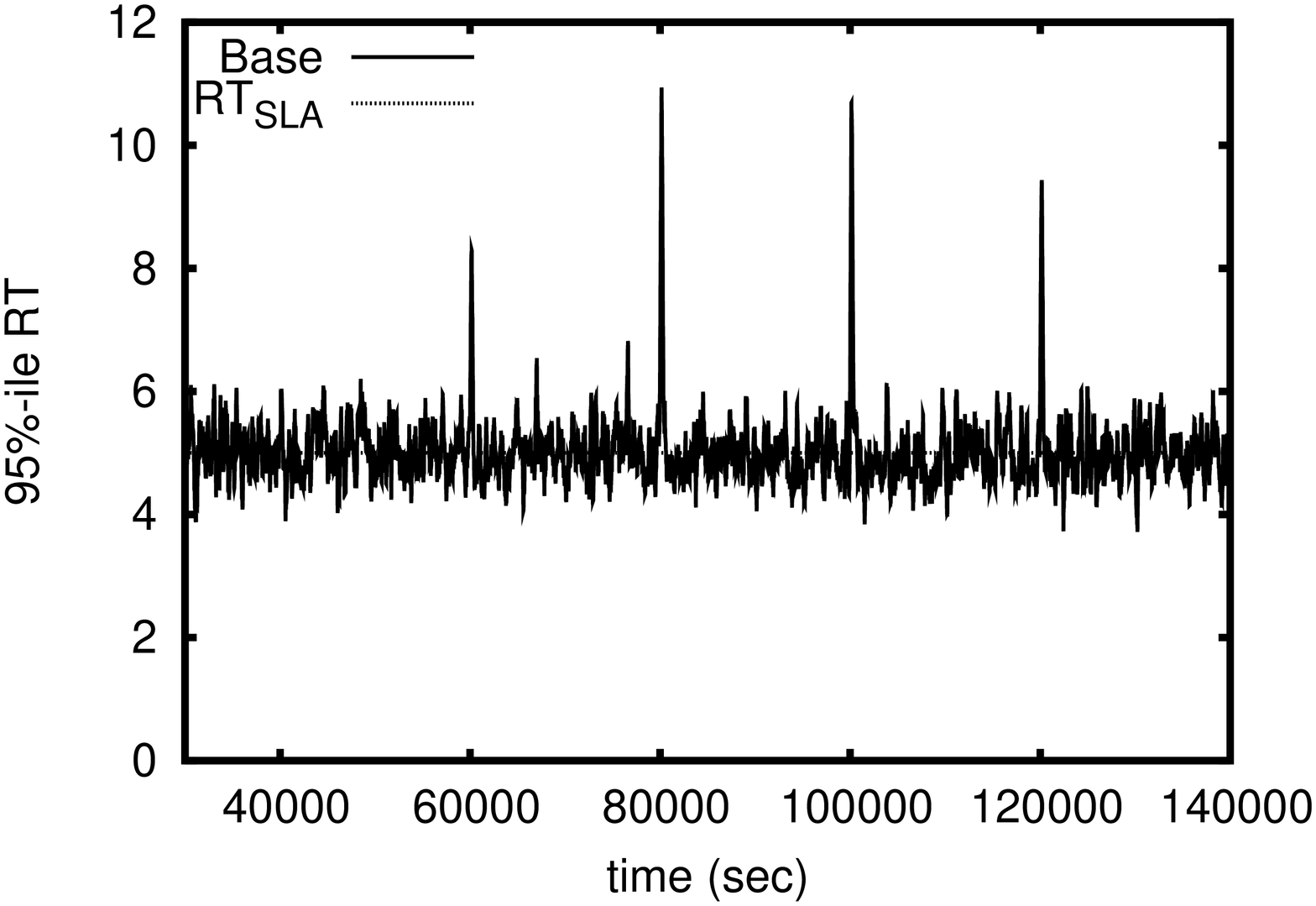}
\caption{95\%-ile (Base)} \label{fig:rt_Base}
\end{figure}

\begin{figure}[!h]
\centering\includegraphics[width = 0.25\textwidth, angle=-90]{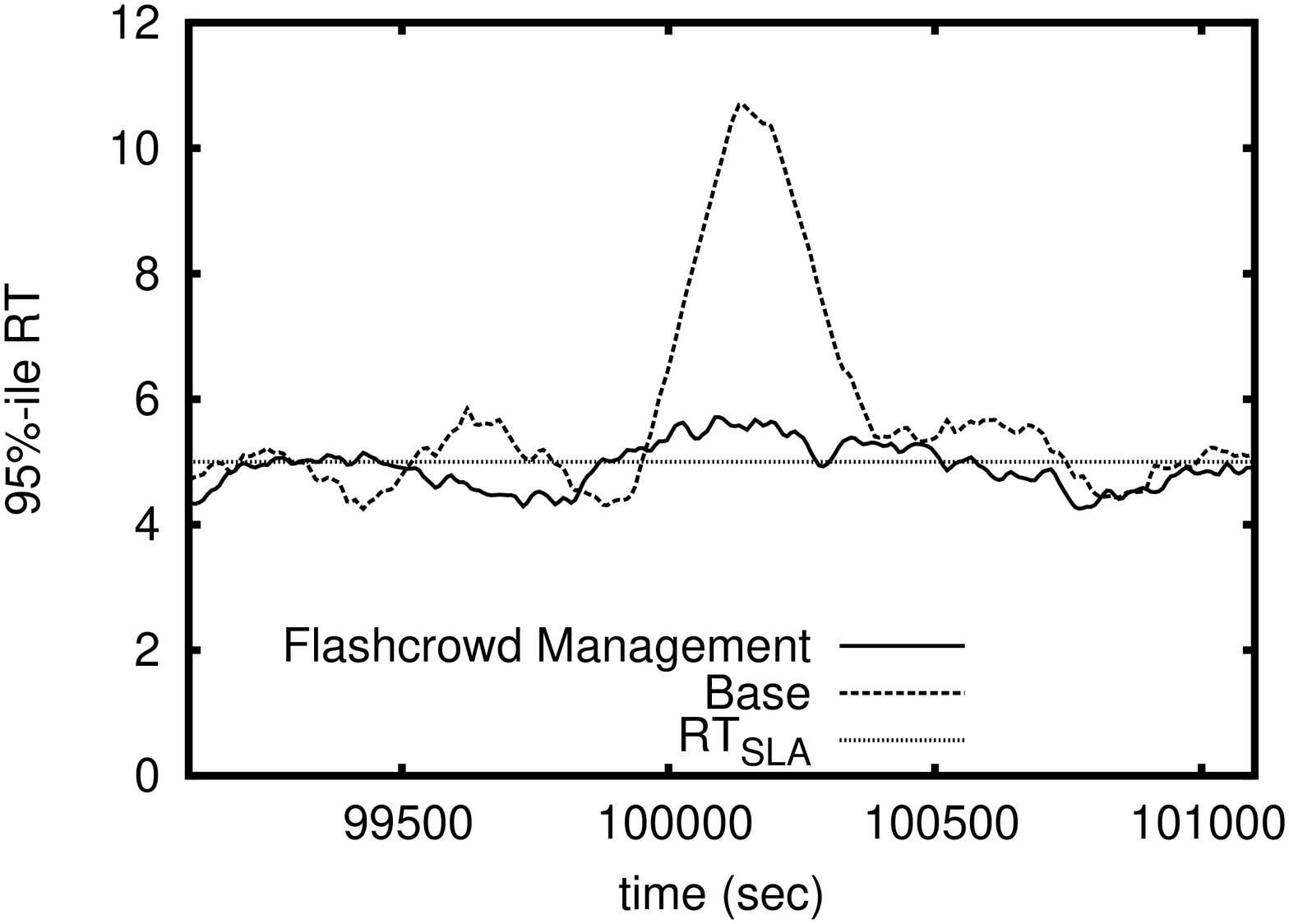}
\caption{95\%-ile RT} \label{fig:zoom_rt_compare}
\end{figure}

\begin{figure}
\centering\includegraphics[width = 0.25\textwidth, angle=-90]{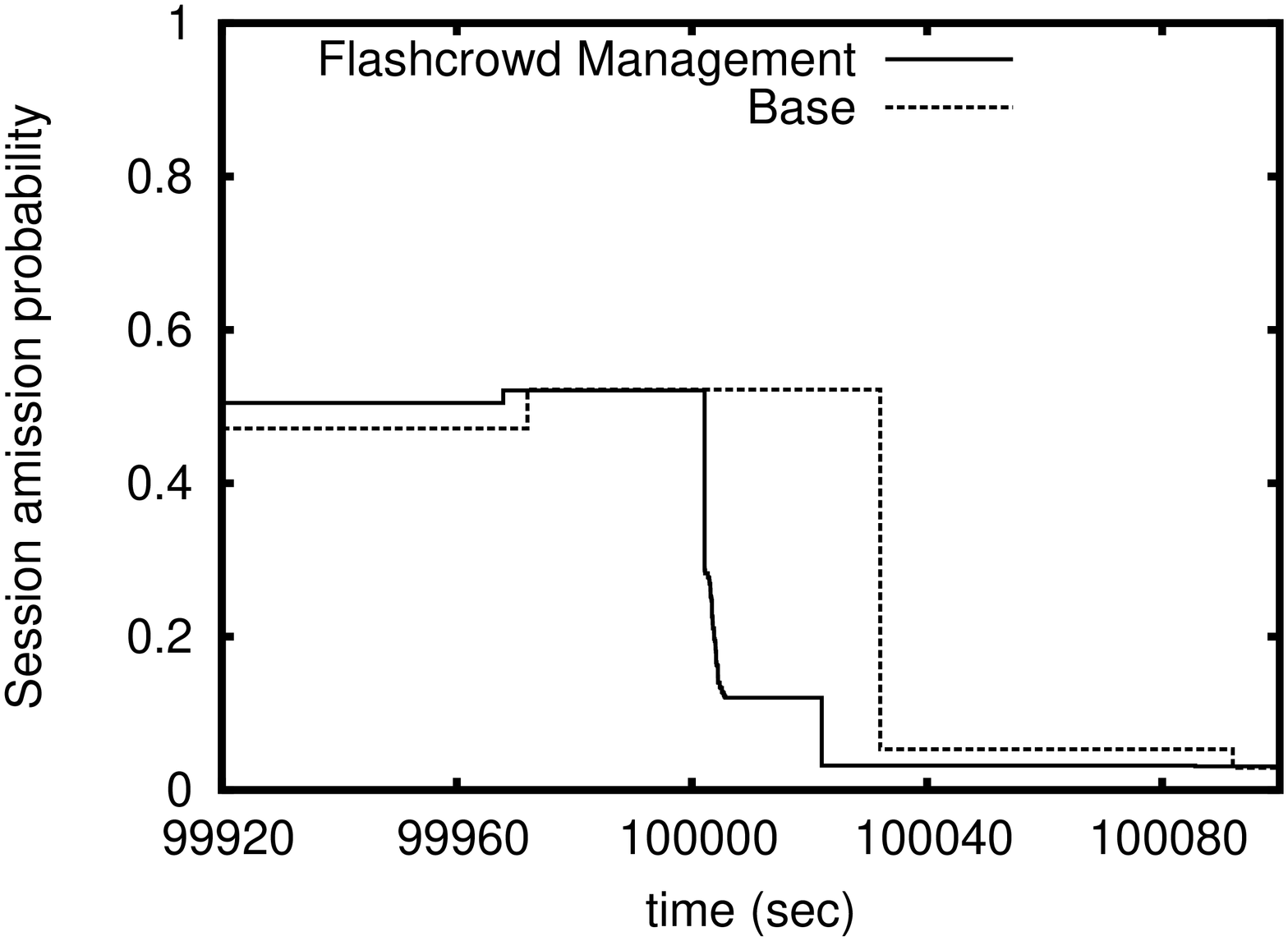}
\caption{Session admission probability} \label{fig:zoom_p_compare}
\end{figure}

In  figures \ref{fig:rt_Base}, \ref{fig:rt_FCM}, \ref{fig:zoom_rt_compare} and   \ref{fig:zoom_p_compare}
 the former version is called {\em Flash Crowd Management} while the latter is called
{\em Base}.
The Base version is the same policy we introduced in
 \cite{Noi_MASCOTS2007} with the addition of the new monitor module detailed in paragraphs \ref{sec:stats_update} and \ref{sec:curve}.

\begin{figure}[!h]
\centering\includegraphics[width = 0.25 \textwidth, angle=-90]{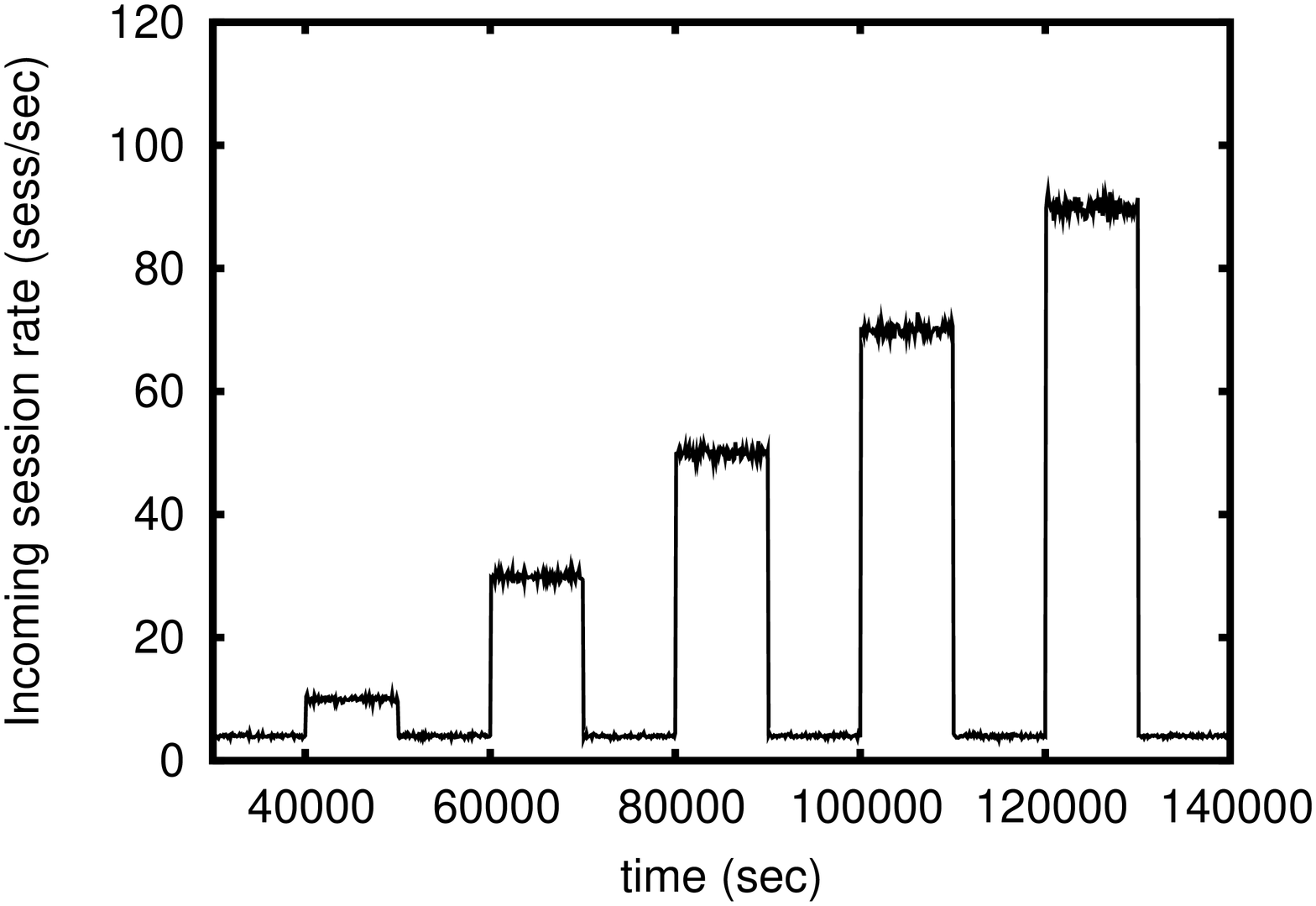}
\caption{Session arrival rate} \label{fig:flash_crowd}
\end{figure}

Figure  \ref{fig:flash_crowd} characterizes the traffic scenario of the last set of experiments. It shows a session arrival rate that is subject to several sudden surges of
growing intensity.

Figures    \ref{fig:rt_Base} and \ref{fig:rt_FCM}
show how the flash crowd management support is capable of
extremely mitigating the spikes of response time caused by the occurrence of flash crowds.
These spikes are instead present in figure \ref{fig:rt_Base} showing that without proper flash crowd management, a violation of the service level agreements is inevitable.


Figures   \ref{fig:zoom_rt_compare} and   \ref{fig:zoom_p_compare} focus on the management of the flash crowd that occurs at 100000 seconds of simulations.

These figures highligth the increased reactivity of SOC when using the
flash crowd management support.
The Base version takes almost 40 seconds to discover the occurrence of the flash crowd and consequently adapt the admission probability, while the enhanced version reacts almost immediately.

Notice the time scale difference between the two figures \ref{fig:zoom_rt_compare} and   \ref{fig:zoom_p_compare}, and the fact that a 40 seconds delay in discovering the flash crowd,  implies the system being in overload for almost 500 seconds.
This is mostly due to the fact that the admission controller works at session granularity.
Notice that premature session interruption would not solve this problem, because on the one hand, sessions are terminating anyway due to client timeout, and on the other hand, the increased session interruption rate should obviously be considered as another aspect of degraded performance.

In particular, figure \ref{fig:zoom_rt_compare} shows how the Base version of SOC is incapable to face such flash crowd, as can be seen by the high values to the 95\%-ile of  response time, which exceed the user time-out. This means that users are abandoning the site due to poor performance or system unavailability.
On the contrary, the flash crowd management enhanced version of SOC is capable of maintaning the response time at acceptable levels by rapidly reducing the session admission probability as soon as the surge in demand is detected.


\section{Related Work}\label{sec:related_work}


There is an impressively growing interest in autonomic computing and
self-managing systems, starting from several industrial initiatives
from IBM \cite{IBM-manifesto}, Hewlett Packard
\cite{HP-design-principles} and Microsoft  \cite{Microsoft_DSI}.
Although self-adaptation capabilities could dramatically improve web system reactivity and
overload control during flash crowds, little effort has been spent on the problem of
autonomous tuning of QoS policies for web systems.



The application of the autonomic computing paradigm to the problem of overload control
in web systems poses some key problems concerning the design of the monitoring module.
The authors of \cite{Pradeep2005} propose a technique for learning
dynamic patterns of web user behavior. A finite state machine
representing the typical user behavior is constructed on the basis
of past history and used for prediction and prefetching techniques.
In paper \cite{Gruser2000} the problem of delay prediction is
analyzed on the basis of a learning activity exploiting passive
measurements of query executions. Such predictive capability is exploited to
enhance traditional query optimizers.

The cited proposals \cite{Gruser2000,Pradeep2005} can partially
contribute to improve the QoS of web systems, but differently from
our work, none of them directly formulate a complete autonomic
solution that at the same time gives directions on how to take
measures, and make corresponding admission control decisions for web
cluster architectures.


The authors of \cite{Liu2005} also address a very important decision
problem in the design of the monitoring module: the timing of
performance control. They propose to adapt the time interval between
successive decisions to the size of workload dependent system
parameters, such as the processor queue length. The dynamic
adjustment of this interval  is of primary importance for threshold
based policies for which a constant time interval between decisions
may lead to an oscillatory behavior in high load scenarios as we show in Section \ref{sec:results}.
Simulations reveal that our  algorithm is not subject to oscillations and shows a very
little dependence on the time interval between decisions.


The problem of designing adaptive component-level thresholds is
analyzed in \cite{Breitgand2005} for a general context of autonomic
computing. The mechanism proposed in the paper consists in
monitoring the threshold values in use by keeping track of false
alarms with respect to possible violations of service level
agreements.
A regression model is used to to fit the observed history.
When a sufficiently confident fit is attained the thresholds are calculated
accordingly. On the contrary if the required confidence is not attained, the thresholds are set to
random values as if there was no history.
 A critical problem of this proposal is the fact that the
most common threshold policies cause on/off behaviors that
often result in unacceptable performance. Our proposal is instead
based on a probabilistic approach and on  a learning technique, that
dynamically creates a knowledge basis for the online evaluation of
the best decision to make even for traffic situations that never
occurred in the past history.


The problem of autonomously configuring a computing cluster to
satisfy SLA requirements is addressed in \cite{Li2005}. This paper
is similar to ours in the design of a strategy for autonomic
computing that divides the problem into different phases, called
{\em monitor}, {\em analyze},  {\em plan} and  {\em execute} (MAPE,
according to the terminology in use by IBM \cite{IBM-MAPE}) in order
to meet SLA requirements in terms of response time and server
utilization. Unlike our work, the authors of this paper designed a
policy whose decisions concern the reconfiguration of resource
allocation to services.

The design of \SOC\ is inspired by the policy AACA we introduced in a previous work \cite{Noi_MASCOTS2007} to which we added the anomaly detection and
decision rate adaptation mechanism that is necessary to
manage flash crowd situations.
With respect to \cite{Noi_MASCOTS2007}, we also largely improved the design of the monitor module as we detail in section \ref{sec:algorithm}.

\section{Conclusion} \label{sec:conclusions}

In this paper we address the problem of overload control for web
based systems. We introduce an original policy, that we name SOC,
 that permits
the self-configuration  and  rapid
adaptivity.
SOC exploits a change detection mechanim to switch between
two modalities according to the time variability of the incoming traffic.

When the incoming traffic is stable, the policy works in normal mode in which
performance controls are paced at a regular rate.
The policy switches to flash crowd management mode as soon as a
rapid surge of demand is detected.
It then increases the rate of performance controls until the incoming traffic becomes more stable.
This permits a fast reaction to sudden changes in traffic intensity, and a high system responsiveness.

Our policy does not require any prior knowledge of the
incoming traffic, nor any assumption on the probability distribution
of request inter-arrival and service time. Unlike other proposals in
the area, our policy works under a wide range of operating
conditions without the need of laborious manual parameter tuning. It
is entirely implemented on dispatching points, without the need of any modification of client
and server software.

We compared our policy to  previously proposed approaches.
Extensive simulations show that it permits an excellent utilization
of system resources while always respecting the limits on response
time imposed by service level agreements. We  show that our
policy reduces the oscillations of response
time common to other policies that
work at session granularity. Simulation results also highlight the
flash crowd management capabilities of SOC, showing how it rapidly adapts
the admission probability to keep the overload under control.

\bibliographystyle{abbrv}
\small{
\bibliography{IWQoS}
}

\end{document}